# Ultra-low loss single-mode silica tapers manufactured by a microheater

Lu Ding, Cherif Belacel, Sara Ducci, Giuseppe Leo, and Ivan Favero

**Using a ceramic thermoelectric heater, we show highly reproducible fabrication of single-mode sub-wavelength silica tapers with ultra-low loss level. The reproducibility of the process is studied statistically, leading to an average taper transmission of 94%. The best tapers have a transmission superior to 99%, above level commonly reached by other fabrication methods. The taper profile is inspected along its length and closely follows the exponential profile predicted by the model of Birks and Li. This high degree of control over the taper shape allows a detailed analysis of the transition to the single-mode regime during tapering. As an application of this fabrication method, we present a micro-looped taper probe for evanescent coupling experiments requiring fine spatial selectivity.**

## I. Introduction

OPTICAL fiber tapers are now versatile components in micro- and nano-optics, with applications in wavelength division multiplexing [1], fiber-based sensors [2] or supercontinuum light generation [3]. After the pioneering work of Knight et al. on a silica microsphere [4], they have also become a popular tool to inject light into micro or nano-cavities by means of evanescent coupling [4]-[8]. For all these applications, a low-loss level is highly desirable, which requires good geometrical regularity of the taper and low surface roughness, both achievable to some extent by pulling the optical fiber in a burner flame. However this technique becomes increasingly difficult and non-reproducible in the case of sub-micron tapers fabrication [9]. Indeed flames are prone to turbulence and can easily fluctuate because of slight bursts in the gas flow or residual air currents in the room. Impure gas can also lead to fiber contamination. By stabilizing the flame and using high purity gas, low linear loss level down to typically $10^{-2}$ dB/mm for 800 nm diameter silica tapers could nevertheless be reached [10]-[12]. A record value of a few $10^{-3}$ dB/mm was reported in [13], corresponding in this case to a taper transmission of 94% when including transition zones. A similar linear loss value of $10^{-3}$ dB/mm was also reported in [14] without notification of the corresponding taper transmission.

The usual flame temperature employed in these silica fiber tapering rigs is 1400-1700°C, close to silica melting point. In the context of taper fabrication, it is however worth noting that at 1100-1200°C, a silica fiber is already soft enough to be stretched slowly. A simple butane burner is hence sufficient for this purpose [11]. This moderate temperature range could also reached using a thermoelectric heater, which has the advantage of providing a more stable heat source. Such heater was for example used in [15] for tapering compound glass nanowires with very low softening temperature of 500°C.

Here we use such micro-heater for tapering Silica fibers, at a temperature of 1160°C. This allows us to circumvent the difficulties inherent to the flame fabrication technique. Thanks to the heat source stability, 800 nm diameter ultra-low loss tapers with transmission of 94% are routinely produced, the best tapers having transmission over 99%. The fabricated tapers are shown to have a well-controlled shape, which allows observing quantitatively their transition to single-mode operation during pulling. We anticipate the high degree of control over taper losses and shape offered by this fabrication method to be beneficial for a broad range of fiber-taper applications. We particularly stress its reproducibility and flexibility. As an example of application, we present a micro-looped single mode taper probe for experiments of evanescent coupling to micro- or nano-cavities.

## II. Taper fabrication and profile analysis

An external cavity laser diode is launched into a single-mode telecom fiber for $\lambda = 1.55$ μm (Corning SMF-28). The fiber coating is striped over 3 cm and the fiber subsequently clamped on two linear translation stages (Newport UTS 100CC) (see Fig. 1a), whose clamps are separated by a distance of 17 cm. The striped part of the fiber is thoroughly cleaned with acetone and isopropanol and positioned in the cavity of a ceramic micro-heater. This micro-heater is a centimeter-sized thermoelectric oven originally developed for fiber-couplers fabrication

[16](see Fig. 1b), and which we operate here at a temperature of 1160°C. After positioning the fiber in the cavity, we wait 120 s to reach thermal equilibrium. The two stages are then set in motion at constant pulling velocity of 40 μm/s, while the fiber transmission is monitored on a photodetector. After about 40 mm of fiber elongation, the stages are stopped and the taper slowly removed from the cavity.

We first mount the produced taper on a glass slide for inspection. Its radius is measured along the fiber axis z with a high magnification ratio optical microscope (500×). The measured radius r(z) is shown in Fig. 2b and compared to the model of Birks and Li, which for a constant hot zone length during tapering predicts [17]:

$$r(z) = \begin{vmatrix} r_0 \exp(\frac{-z - z_0 - L_0/2}{L_0}) & -z_0 - L_0/2 \leq z \leq -L_0/2 \\ r_w = r_0 \exp(\frac{-z_0}{L_0}) & -L_0/2 \leq z \leq L_0/2 \\ r_0 \exp(\frac{z - z_0 - L_0/2}{L_0}) & L_0/2 \leq z \leq z_0 + L_0/2 \end{vmatrix} \quad (1)$$

where $L_0$ is the hot zone length [17], $2z_0$ the elongation length of the fiber and $z = 0$ the middle of the taper. Although the temperature profile of the heater hot zone is closer to Gaussian than to a constant flat profile [16], Eq.(1) describes well our results. The fit of the measured profile is excellent when using the measured values $z_0$=19.75 mm and $r_0$=66 μm, and setting the only adjustable parameter $L_0$ to 3.96 mm. On the taper edges, we note a small discrepancy due to temperature conditions during early pulling which deviate from subsequent steady-state conditions. However, the data show convincingly the validity of the model down to the scale of sub-micron tapered fibers, whereas it was originally only tested on much larger glass rods of diameter 5 mm [17]. Here the waist diameter is 0.9 ± 0.3 μm. In our fabrication method, we can also vary $L_0$ by changing the operation temperature of the heater and the pulling velocity. This allows fabricating tapers of various profiles, depending on the targeted mechanical or optical application.

### III. HIGHLY REPRODUCIBLE ULTRA-LOW LOSS LEVEL

The profile measured in Fig. 2 is expected to be highly adiabatic for light at λ = 1.55 μm. Here by using the thermo-electric oven, we indeed easily obtain highly adiabatic tapers, which reproducibly show extremely low-loss level. In Fig. 3, we present a statistical study performed over 80 tapers, all having about 40 mm of elongation length. The produced tapers have an average transmission level of 94 %, on par with the best sub-wavelength tapers previously reported [10]-[13]. But here 94 % is an average result and we stress that half of the tapers display in reality a transmission comprised between 95 and 100 %. The best tapers rise to transmission values superior to 99 %, that we do not further resolve. If including taper transitions losses, this corresponds for the uniform waist region to a linear loss level in the $10^{-3}$ dB/mm range. Similar ultra-low values were inferred in [13,14]. However, for what concerns the evaluation of linear loss at the waist, the contribution of taper transitions losses precludes a rigorous comparison with our current results. Only an upper bound for the linear loss at the waist can be given. For what concerns the total taper transmission, our fabrication method gives reproducibly excellent results, above the level commonly reached with other methods.

In ambient conditions, the taper transmission decays upon time because of silica reaction with water molecules in the air [18]. We measured the decay rate for various relative humidity (RH) levels at a temperature of 25°C (Table 1). In a wet environment (RH=35 %), the taper lifetime can be as short as a few hours. In a plexiglass enclosure purged with dried air, we observe taper transmission constant over weeks.

### IV. TRANSITION TO A SINGLE MODE TAPER

On top of achieving low-loss level, it is in many applications required to produce tapers that are single mode guiding. This is the case in optics experiments where ideality relies on perfect mode-to-mode coupling or in filtering applications, where higher order modes need be suppressed. Following the cut-off criterion for a weakly guiding taper [19], the taper becomes single-mode when its waist radius $r_w$ is smaller than $(2.405\lambda/2\pi NA)$, where NA is the numerical aperture of the fiber NA=$(n_{co}^2-n_{cl}^2)^{1/2}$ with $n_{co}$ and $n_{cl}$ the refractive of the core and cladding respectively. For $n_{cl}=1$ (air) and $n_{co}=1.45$ at $\lambda=1.55$ µm, this occurs for a waist diameter inferior to 1.13 µm. The weak guidance approximation ($\Delta<<1$) is sufficient for our purpose here but we note that a fiber taper is not a very weakly guiding structure at its waist since its profile height parameter $\Delta=(n_{co}^2-n_{cl}^2)/2n_{co}^2$ amounts to 0.26, when standard telecom fibers have $\Delta$ of order 0.001 [20]. Upon stretching, the fiber transmission as a function of elongation length $2z_0$ displays oscillations of varying amplitude and frequencies (see Fig. 4 upper panel). These oscillations arise from interferences between different modes supported by the fiber taper and their vanishing means that the single-mode operation is reached. However as we will see below in a detailed study, this criterion must be used with some care in the experiments.

These oscillations of the transmission can be analysed using the mathematical tool of the Gabor transform, which is a specific form of short-time Fourier analysis [21]. Fig. 4 shows a taper transmission as function of the elongation (upper panel) and its Gabor transform (lower panel). The latter has a dominant component vanishing at elongation length of 32 mm, exactly when a dramatic change in the beating amplitude occurs. A second component of smaller weight vanishes later at a pulling length of about 34 mm. To model this behavior, we use analytical approximate expressions (accurate far from cut-off) for the propagation constants $\beta$ of modes guided in a dielectric cylinder of radius r [22]:

$$\left| \begin{array}{l} \beta(r) = \left[\left(\frac{2\pi}{\lambda}\right)^2 n_{co}^2 - \left(\frac{U(r)}{r}\right)^2\right]^{1/2} \\ U(r) = U(\infty) \times \exp\left(\frac{-\lambda r}{2\pi NA}\right) \end{array} \right.$$

where $U(\infty)=2.405, 3.832, 5.135, 5.520$ for the four first distinct propagation constants $\beta_{k=1,2,3,4}$ of guided modes that we consider here. For each value of $2z_0$ we express the phase accumulated by a mode of propagation constant $\beta_k$ during transmission through the taper [11]:

$$\Phi_k(2z_0) = \beta_k(r_w)L_0 + 2\int_0^{z_0} \beta_k(r(z))\,dz$$

where $r_w$ and $r(z)$ given by Eq. (1). To model the taper transmission T, we take the modulus square of a weighted sum of the $\exp(i\phi_k)$, where the weights are adjusted to reproduce the transmission beating amplitude (not shown here). The white dashed lines in the lower panel are obtained by a Gabor transform of the obtained T, where the only adjustable parameter $L_0$ is set to 3.795 mm. The close agreement with data allows us to interpret the dominant component in the Gabor transform as originating from a beating of the taper fundamental mode $HE_{11}$ with the $HE_{12}$ mode of propagation constant $\beta_4$. The second fainter component relates to a beating with other modes of lower symmetry analogous to $LP_{11}$. Note that complete modeling close to cut-off is also possible if solving numerically the Helmholtz equation [11]. Fig. 4 shows that a detailed analysis is absolutely needed to ensure that the taper has indeed reached the single-mode regime. If simply stopping the pulling when the dominant beating disappears at 32 mm of elongation, the obtained taper still supports three guided modes $HE_{11}$, $TE_{01}$ and $HE_{21}$.

## V. MICRO-LOOPED FIBER TAPER

Straight optical fiber tapers are employed for evanescent coupling to small optical cavities like silica microspheres or microtoroids [4], [5] but in the case of sub-micron cavities integrated on-chip, parasitic coupling to the substrate becomes limiting with a straight taper. To circumvent this difficulty, various strategies have been devised to locally curve the taper on a distance of about 100 µm to improve the spatial selectivity of the coupling [6-8,12]. Here building on an earlier method [8], we present and characterize a simple technique for fabricating a micro- looped fiber taper for spatially selective evanescent coupling applications.

In our fabrication process, after removing the taper from the micro-heater, the translation stages are both moved 1 mm back in order to release the taper tension. Using a rotating fiber clamp, the taper is first twisted 4 turns on itself with the effect of forming a multi-twisted looped structure at its center as shown in Fig. 5a. The taper is put back under tension moving both stages by 0.3 mm and glued on one end on a acrylate plate with epoxy. Its opposite end is then pulled gently while observing the loop un-twist till it reaches the single loop structure shown in Fig. 5b. It is eventually glued on the plate and the exact degree of tension adjusted during curing. The looped tapers obtained in this manner have a typical diameter of 70 µm (Fig. 5b inset) and transmission spectrum shown in Fig. 5c. Periodic oscillations in this spectrum correspond to resonances of the micro-ring cavity formed by the loop [23], with a free spectral range of 8.2 nm. The taper waist of diameter 1 µm having an effective index of 1.25 for its single guided mode at λ = 1.55 µm, it corresponds to a loop diameter of 74 µm, in good agreement with the microscope observation. The free spectral range measurement provides an in-situ control of the loop geometry, a beneficial feature for adjusting its mechanical properties in evanescent coupling experiments. Using this adjustable fiber taper structure, we have recently probed optical gallery modes of GaAs disks of a few µm diameter and 200 nm thickness (to be described elsewhere).

## VI. CONCLUSION

Given the recent success of sub-wavelength optical fiber taper in the arena of micro and nano-photonics, it is desirable to increase access to these techniques in optics laboratories. The fabrication method used in this article is highly reproducible, low-cost and easy to implement. It circumvents difficulties of the standard flame fabrication and offers ultra-low loss level tapers on demand, together with a high level of control over their shape and single-mode operation. Further progress on the way to large-scale applications could include taper encapsulation for use in ambient conditions.

## ACKNOWLEDGMENT

We thank Professor Hiromasa Ito and Pascale Senellart for fruitful discussion.

## REFERENCES

[1] F. Gonthier, S. Lacroix, X. Daxhelet, R. J. Black, and J. Bures, "Broad-band all-fiber filters for wavelength division multiplexing application," Appl. Phys. Lett. **54**, 1290 (1989).
[2] L. C. Boob, and P. M. Shankar, "Tapered optical fiber components and sensors," Microwave J. **35**, 218 (1992).
[3] T. A. Birks, W. J. Wadsworth, and P. S. J. Russel, "Supercontinuum generation in tapered fibers," Opt. Lett. **25**, 1415 (2000).
[4] J. C. Knight, G. Cheung, F. Jacques, and T. Birks, "Phase-matched excitation of whispering gallery mode resonances by a fiber taper," Opt. Lett. **22**, 1129 (1997).
[5] M. Cai, and K. Vahala, "Highly efficient optical power transfer to whispering gallery modes by use of a symmetrical dual coupling configuration," Opt. Lett. **25**, 260 (2000).
[6] K. Srinivasan, P. E. Barclay, M. Borselli, and O. Painter, "Optical fiber based measurement of an ultrasmall volume, high-Q photonic crystal microcavity," Phys. Rev. B **70**, 081306 (2004).
[7] I. K. Hwang, S. K. Kim, J. K. yang, S. H. Kim, S. H. Lee, and Y. H. Lee, "Curved microfiber photon coupling for photonic crystal light emitter," Appl. Phys. Lett. **87**, 131107 (2005).
[8] C. Grillet, C. Monat, C. L. C. Smith, B. Eggleton, D. J. Moss, S. Frederick, D. Dalacu, P. J. Poole, J. Lapointe, G. Aers, and R. L. Williams, "Nanowire coupling to photonic crystal nanocavities for single photon sources," Opt. Exp. **15**, 1267 (2007).
[9] L. Tong, R. R. Gattas, J. B. Ashcom, S. He, J. Lou, M. Shen, I. Maxwell, and E. Mazur, "Sub-wavelength diameter silica wires for low-loss optical wave guiding," Nature **426**, 816 (2003).
[10] G. Brambilla, V. Finazzi, and D. J. Richardson, "Ultra-low-loss optical fiber nanotapers," Opt. Exp. **12**, 2258 (2004).
[11] F. Orucevic, V. Lefevre-Seguin, and J. Hare, "Transmittance and near field characterization of sub-wavelength tapered optical fibers," Opt. Exp. **15**, 13624 (2007).
[12] C. P. Michael, M. Borselli, T. J. Johnson, C. Chrystal, and O. Painter, "An optical fiber taper probe for wafer scale microphotonics device characterization," Opt. Exp. **15**, 4745 (2007).
[13] S. G. Leon-Saval, T. A. Birks, W. J. Wadsworth, P. St. J. Russel, and M. W. Mason, "Supercontinuum generation in submicron fibre waveguides," Opt. Exp. **12**, 2864 (2004).
[14] G. Brambilla, F. Xu, and X. Feng, "Fabrication of optical fibre nanowires and their optical and mechanical characterisation", Electronics Lett. **42** (9), 517 (2006).
[15] G. Brambilla, F. Koizumi, X. Feng, and D. J. Richardson, "Compound-glass optical nanowires," Electronics Lett. **41** (7), 400 (2005).
[16] Y. Takeuchi, and J. Noda, "Novel fiber coupler tapering process using a microheater," IEEE Phot. Tech. Lett. **4**, 1041 (1992).
[17] T. A. Birks, and Y. W. Li, "The shape of fiber tapers," IEEE J. Light. Tech. **4**, 1041 (1992).
[18] J. L. Mrotek, M. J. Matthewson, C. R. Kurkjian, "The fatigue of high strength fused silica optical fibers in low humidity," J. Non-crystalline. Solids. **297**, 91 (2002).
[19] B. E. A. Saleh, and M. C. Teich, *Fundamentals of Photonics* (Wiley, 2$^{nd}$ed, 2007).
[20] A. W. Snyder, and J. D. Love, *Optical waveguide theory* (Chapmann ad Hall, London, 1983).
[21] http://en.wikipedia.org/wiki/Gabor_transform
[22] A. W. Snyder, "Asymptotic expressions for eigenfunctions and eigenvalues of a dielectric or optical waveguide," IEEE Transactions on Microwave Theory and Techniques **17**, 1130 (1969).
[23] M. Sumetsky, Y. Dulashko, and A. Hale, "Fabrication and study of bent and coiled free silica nanowires: self-coupling microloop optical interferometer", Opt. Exp. **12**, 3522 (2004).

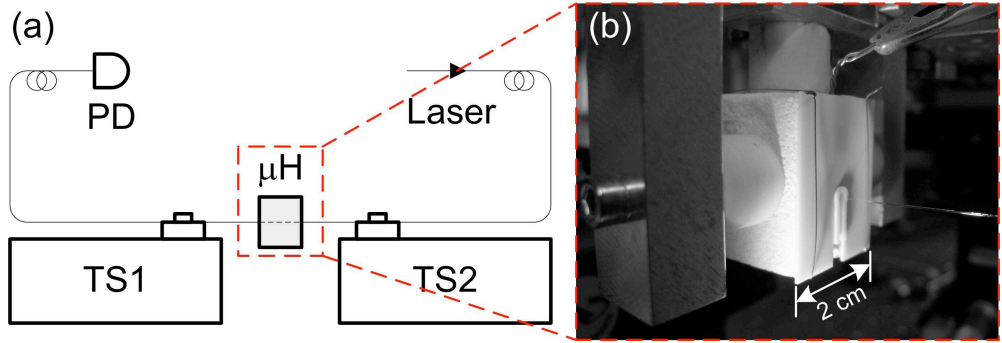

Fig. 1. Micro-heater taper fabrication. (a) Schematic view of the set-up for fiber taper fabrication. µH is the thermoelectric micro-heater providing a stable hot zone. TS 1,2 are motorized linear translation stages. PD is the photodetector. (b) Zoom-in picture of the micro-heater with the fiber positioned in the cavity.

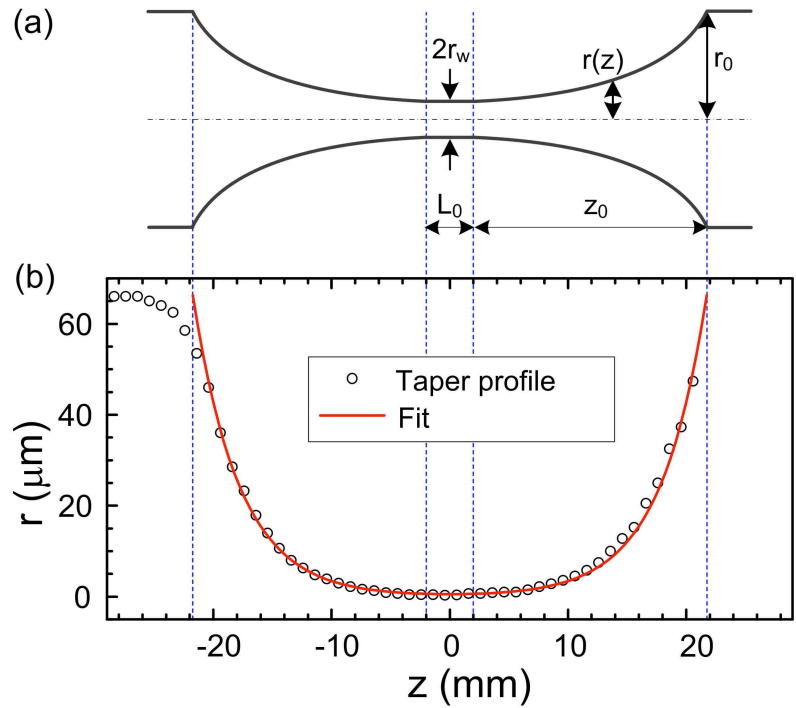

Fig. 2. Taper profile. (a) Schematic profile of a symmetric fiber taper formed in a constant hot zone, where the total taper length is $2z_0+L_0$. $z_0$ is the length of the tapering transition and $L_0$ the length of the taper waist (also the hot zone size here). (b) Measured profile of a typical fiber taper fabricated using the micro-heater. Here the taper waist diameter is 0.9 µm.

TABLE I
TRANSMISSION DECAY OF FIBER TAPER TRANSMISSION (IN UNITS OF %/HOUR) WITH RESPECT TO THE RELATIVE HUMIDITY OF THE ENVIRONMENT AT 25ºC.

| Relative humidity (%) | Transmission decay (%/hour) |
|---|---|
| < 20 | 0.2 |
| 25 | 9 |
| 30 | 17 |
| 34 | 30 |

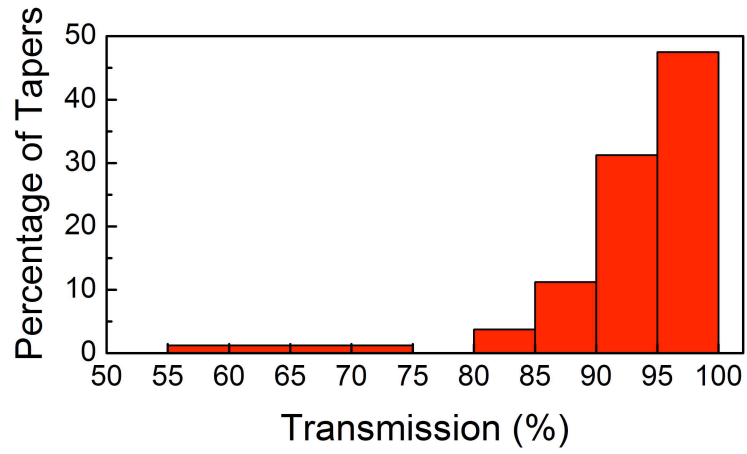

Fig. 3. Reproducible ultra-low loss level. Statistic histogram of fiber tapers transmission out of 80 samples.

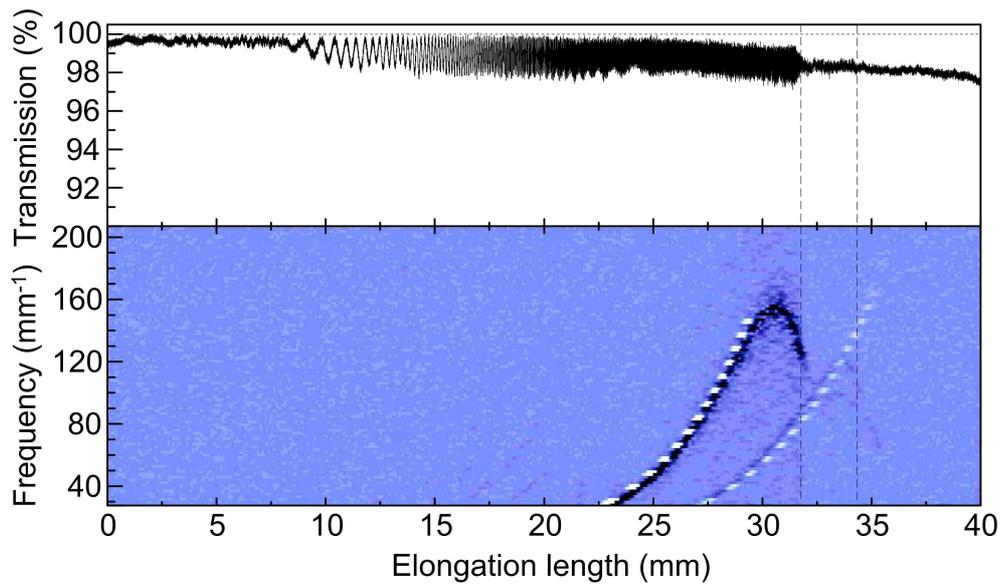

Fig. 4. Transition to single-mode operation. Upper panel: recorded fiber taper transmission as a function of the elongation length. Lower panel: Gabor transform of the recorded transmission. Dashed lines are given by our model.

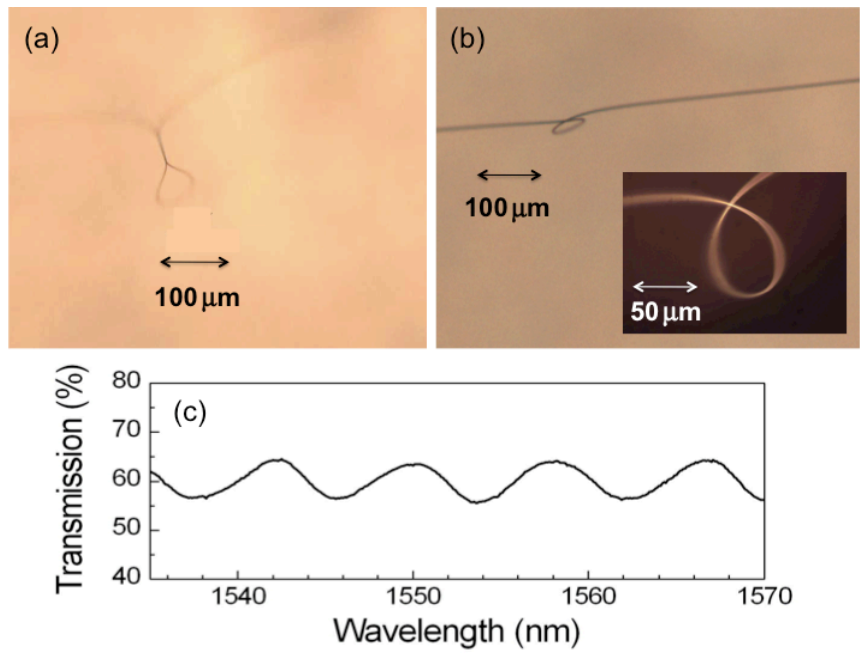

Fig. 5. Micro-looped taper. (a) Multi-twisted looped taper structure. (b) Tensioned single loop taper. (c) Transmission spectrum of a single loop taper.